# TIME VARIATIONS OF THE SUPERKAMIOKANDE SOLAR NEUTRINO FLUX DATA

Abu Salem Mandal[1], Koushik Ghosh[2] and Probhas Raychaudhuri[1]

[1]Department of Applied Mathematics
University of Calcutta
92, A.P.C. Road, Calcutta- 700 009
INDIA

[2]Department of Mathematics
Dr. B.C. Roy Engineering College
Durgapur – 713 206
INDIA

**Abstract:**

We have used the Date-Compensated Discrete Fourier Transform and Periodogram analysis of the solar neutrino flux data from 1) 5-day-long samples from Super-Kamiokande-I detector during the period from June, 1996 to July, 2001; 2) 10-day-long samples from the same detector during the same period and 3) 45-day-long samples from the same detector during the same period. (1) exhibits periodicity around 0.21-0.22, 0.67-0.77, 1.15-1.98, 6.72-6.95, 12.05-13.24, 22.48-24.02, 33.50 and 40.73 months. (2) shows periodicity around 0.39-0.45, 1.31-2.23, 5.20-5.32, 9.43-11.65, 13.54-14.38, 24.54, 32.99 and 41.69 months. For (3) we observe periodicity around 1.61, 14.01, 24.06, 32.50 and 42.03 months. We have found almost similar periods in the solar flares, sunspot data, solar proton data ( $\bar{\epsilon}$ >10 Mev) which indicates that the solar activity cycle may be due to the variable character of nuclear energy generation inside the sun.

**PACS number(s):** 14.60.Pq, 26.65.+t, 95.85.Ry, 96.40.Tv

## I. Introduction:

Solar neutrino flux detection is very important not only to understand the stellar evolution but also to understand the origin of the solar activity cycle. Recent solar neutrino flux observed by Super-Kamiokande [1] and SNO detectors[2] suggest that solar neutrino flux from $^8B$ neutrino and $^3He+p$ neutrino from Standard Solar Model (S.S.M.)[3] is at best compatible with S.S.M. calculation if we consider the neutrino oscillation of M.S.W.[4] or if the neutrino flux from the sun is a mixture of two kinds of neutrino i.e. $\nu_e$ and $\nu_\mu$ [5] . Standard Solar Model (S.S.M.) are known to yield the stellar structure to a very good degree of precision but the S.S.M. cannot explain the solar activity cycle, the reason being that this S.S.M. does not include temperature and magnetic variability of the solar core[6,7]. The temperature variability implied a variation of the energy source and from that source of energy magnetic field can be generated which also imply a magnetic variability [7]. The temperature variation is important for the time variation of the solar neutrino flux. So we need a perturbed solar model and it is outlined by Raychaudhuri since 1971[6,7], which may satisfy all the requirements of solar activity cycle with S.S.M.. For the support of perturbed solar model we have demonstrated that solar neutrino flux data are fractal in nature [8]. The excess nuclear energy from the perturbed nature of the solar model transformed into magnetic energy, gravitational energy and thermal energy etc. below the tachocline. The variable nature of magnetic energy induces dynamic action for the generation of solar magnetic field.

Recently Yoo *et al.* (2003)[9] searched the periodic modulations of the solar neutrino flux data of Super-Kamiokande-I (S.K.-I) detector from 31 May 1996 to 15 July 2001, almost half of the solar activity cycle, yielding a total detector life time of 1496 days. The solar neutrino data from S.K., acquired for 1871 elapsed days from the beginning of data are divided into roughly 10-day-long samples as listed in table I of

Yoo *et al.*[9]. It is observed that not all of the data are perfectly of 10 days. They used Lomb periodogram method for unevenly arranged sample data to search for possible periodicities in the S.K.-I solar neutrino flux data. They have found no statistical significance of the periodicities in the S.K.-I solar neutrino flux data. However, Caldwell and Sturrock[10] used almost the same method i.e. Lomb-Scargle method of analysis and they have found a very interesting period of 13.75 days in the solar neutrino flux data of S.K.-I apart from other periods. Thus there arises a controversy regarding the periodicities of the solar neutrino flux data. Raychaudhuri [11] analysed the solar neutrino flux data of S.K.-I 45-days-sample data and have found 5 and 10 months period in the data and the same periods are also found in the $^{37}$Cl, SAGE and GALLEX solar neutrino flux data. 5 months period is seen in many solar activities (e.g. solar flares, sunspot etc.) indicating a relation between solar internal activities and solar surface activities.

The purpose of the paper is to see whether the Super-Kamiokande-I solar neutrino flux data is variable in nature or not. The observation of a variable nature of solar neutrino would provide significance to our understanding of solar internal dynamics and probably to the requirement of the modification of the Standard Solar Model i.e. a perturbed solar model. In this paper we shall study the solar neutrino flux data from S.K.-I 5-days-sample data, 10-days sample data and 45-days-sample data during the period from 31 May 1996 to 15 July 2001. We shall first study whether the data samples given by S.K.-I collaborators are random in nature or not. If they are random then there may not be the possibility of any distribution of the data or any periodicities in the data of S.K.-I. If the data are non-random in nature then there is a possibility of periodicity in the S.K.-I data.

## II. Statistical Test of Randomness:

It is observed that S.K.-I solar neutrino flux data from 31 May 1996 to 15 July 2001 for 5-days, 10-days, 45-days data samples have large statistical errors from 15% to almost 30%. It is very difficult to evaluate precisely the statistical analysis of all the data. Without filtering we first evaluate the randomness of the data [12] from where Caldwell and Sturrock [10] confirmed their periodicities and Yoo *et al.*[9] have not found the periodicities in the S.K.-I data. We use the run test for the evaluation of data without filtering.

We have found that 5-days, 10-days and 45-days data of S.K.-I are random in nature while 30-days data evaluated from the 45-days data are not random. It is expected that 30-days data may have a period and we have found 5-months period with 99% confidence level.

It is already mentioned that original data of S.K.-I have errors 15% to 30%. So, it is necessary to smooth the data by filtering. The simplest filtering is the moving average method. So, we use 3 point moving average of the 5-days, 10-days and 45-days data of S.K.-I.

After making the moving average we have seen that all the data are non-random in nature and so the moving average data must follow a distribution. Hence the obtained moving average data can be satisfactorily used for the analysis of periodicity following the methods 1) Ferraz-Mello (1981) [13] method of Date-Compensated Discrete Fourier Transform and 2) Scargle (1982) [14] method of Periodogram.

## III. Calculation of Periodicities:

### 1. Ferraz-Mello Method of Date-Compensated Discrete Fourier Transform:

The technique called Date-Compensated Discrete Fourier Transform (DCDFT) corresponds to a curve-fitting approach using a sinusoid-plus-constant model and is summarized below. For each trial frequency ω, one coefficient of spectral correlation S is obtained by the following formulae [13].

$$a_0^{-2} = N \tag{1}$$

$$a_1^{-2} = \sum \cos^2 x_i - a_0^2 \left(\sum \cos x_i\right)^2 \tag{2}$$

$$a_2^{-2} = \sum \sin^2 x_i - a_0^2 \left(\sum \sin x_i\right)^2 - a_1^2 M^2 \tag{3}$$

where

$$M = \sum \cos x_i \sin x_i - a_0^2 \left(\sum \sin x_i\right) \left(\sum \cos x_i\right) \tag{4}$$

and

$$C_1 = a_1 \sum f_i \cos x_i \tag{5}$$

$$C_2 = a_2 \sum f_i \sin x_i - a_1 a_2 C_1 M \tag{6}$$

$$S = \frac{C_1^2 + C_2^2}{\sum f_i^2} \tag{7}$$

N is the number of observations in the series, $t_i$ are the observation dates, $x_i = 2\pi\omega t_i$ and $f_i$ are the measures referred to the mean i.e. $f_i = x_i - \bar{x}$ so that $\sum f_i = 0$. All other symbols are intrinsic quantities. The summations are made for i=1 to i=N. Usually the range of frequencies is considered from 0 to the Nyquist frequency and the Nyquist frequency is given by $\omega_N = (N/2)\omega_{obs}$ where $\omega_{obs} = 2\pi/T$ i.e. $\omega_N = \pi N/T$.

To decide whether the peaks in the graph are significant or not we use the test derived by G.R. Quast [15] which is given by the following expressions:

$$G = -\frac{N-3}{2}\ln(1-S) \tag{8}$$

$$H = \frac{N-4}{N-3}(G + e^{-G} - 1) \tag{9}$$

$$\alpha = \frac{2(N-3)\Delta t . \Delta \omega}{3(N-4)} \qquad (10)$$

$$C = (1 - e^{-H})^{\alpha} \qquad (11)$$

where $\Delta T$ is the time interval covered by the observations and $\Delta\omega$ is the range of frequencies sampled. C is the confidence of the result. (1−C) may be interpreted as the probability of having the height of the highest peak by chance only. Here we consider 500 values of $\omega$ from $\omega_{obs}$ = 1/N to $\omega_N$ and we arrange the corresponding months from 1/$\omega_N$ to 1/$\omega_{obs}$ at equal intervals and we obtain corresponding magnitudes of H and C at those months.

**2.     Scargle Method of Periodogram:**

For a time series X(t), where i = 1, 2, …, N the periodogram as a function of the frequency $\omega$ is defined as[14] :

$$P_X(\omega) = \frac{1}{2}\left\{ \frac{\left[\sum_{i=1}^{N} X(t_i)\cos\omega(t_i - \tau)\right]^2}{\sum_{i=1}^{N} \cos^2\omega(t_i - \tau)} + \frac{\left[\sum_{i=1}^{N} X(t_i)\sin\omega(t_i - \tau)\right]^2}{\sum_{i=1}^{N} \sin^2\omega(t_i - \tau)} \right\} \qquad (12)$$

where $\tau$ is defined by the equation

$$\tan(2\omega\tau) = (\sum_{i=1}^{N} \sin 2\omega t_i)/(\sum_{i=1}^{N} \cos 2\omega t_i) \qquad (13)$$

Here also we consider 500 values of $\omega$ from $\omega_{obs}$ = 1/N to $\omega_N$ and we arrange the corresponding months from 1/$\omega_N$ to 1/$\omega_{obs}$ at equal intervals and we obtain corresponding $P_X(\omega)$ at those months.

## IV. Results:

| Data | Periods (in months) | |
|---|---|---|
| | **Ferraz-Mello Method of Date-Compensated Discrete Fourier Transform** | **Scargle Method of Periodogram** |
| 1) 5-day-long samples from Super-Kamiokande-I detector during the period from June, 1996 to July, 2001 | 0.21 (79.85%), 0.77 (79.85%), 1.98 (99.79%), 3.40 (99.73%), 6.72 (99.92%), 12.05 (99.99%), 22.48 (99.81%), 33.50 (99.99%). | 0.22, 0.33, 0.67, 1.15, 6.95, 9.21, 13.24, 24.02, 40.73. |
| 2) 10-day-long samples from Super-Kamiokande-I detector during the period from June, 1996 to July, 2001 | 0.45 (88.24%), 1.31 (88.30%), 5.20 (99.92%), 11.65 (99.96%), 14.38 (99.99%), 32.99 (99.99%). | 0.39, 2.23, 5.32, 9.43, 13.54, 24.54, 41.69. |
| 3) 45-day-long samples from Super-Kamiokande-I detector during the period from June, 1996 to July, 2001 | 14.01 (98.17%), 32.50 (99.99%). | 1.61, 24.06, 42.03. |

**N. B.** The percentage values within bracket give the corresponding confidence levels.

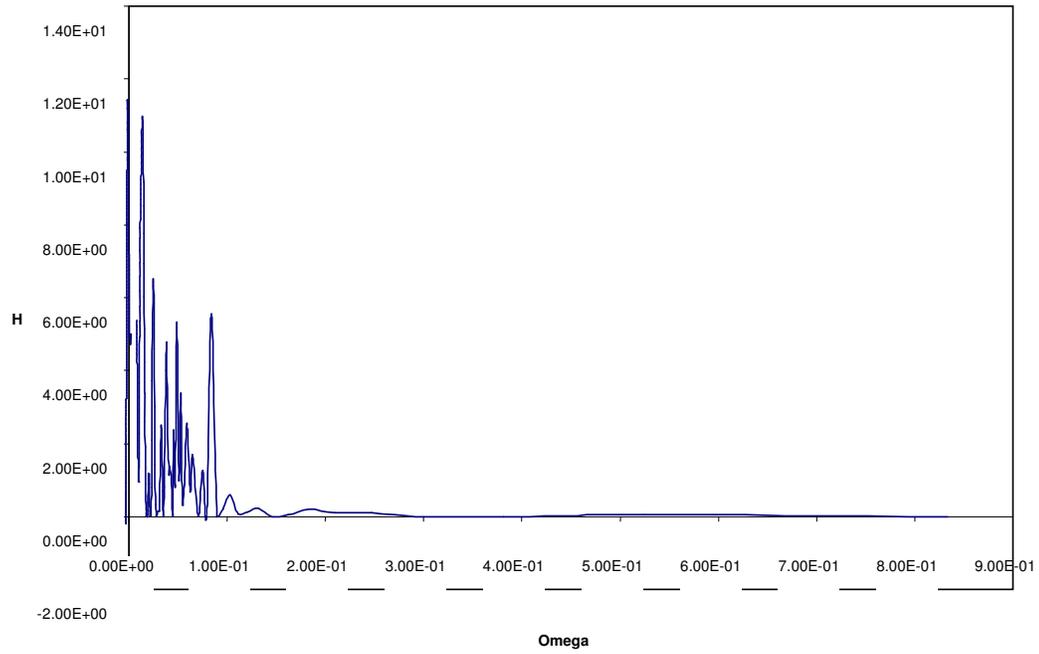

**Fig. 1:** H-ω graph obtained by Ferraz-Mello method of 5-day-long samples from Super-Kamiokande-I detector during the period from June, 1996 to July, 2001.

**Fig. 2:** H-ω graph obtained by Ferraz-Mello method of 10-day-long samples from Super-Kamiokande-I detector during the period from June, 1996 to July, 2001.

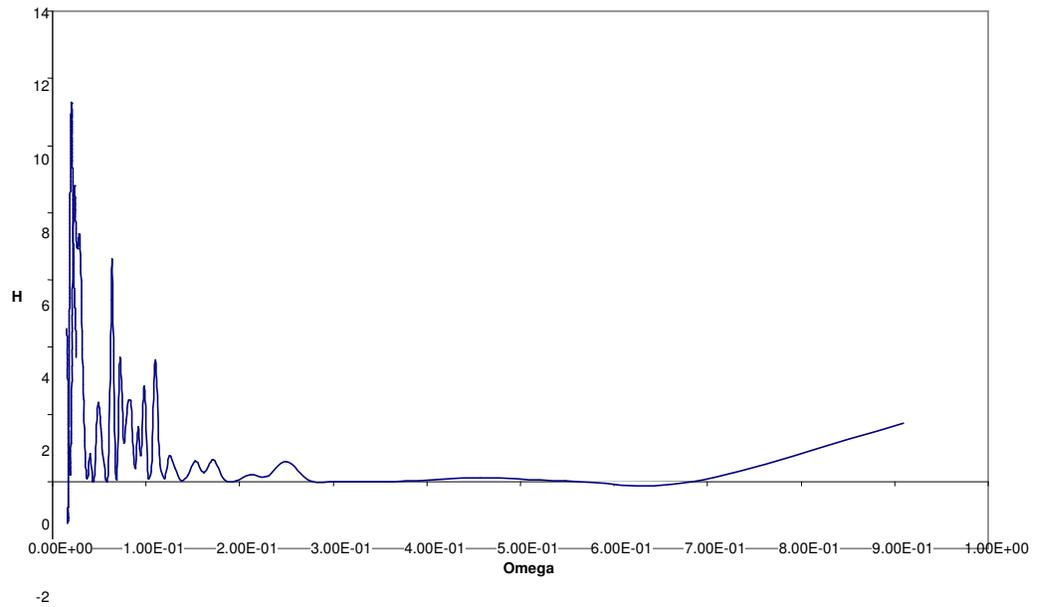

**Fig. 3:** H-ω graph obtained by Ferraz-Mello method of 45-day-long samples from Super-Kamiokande-I detector during the period from June, 1996 to July, 2001.

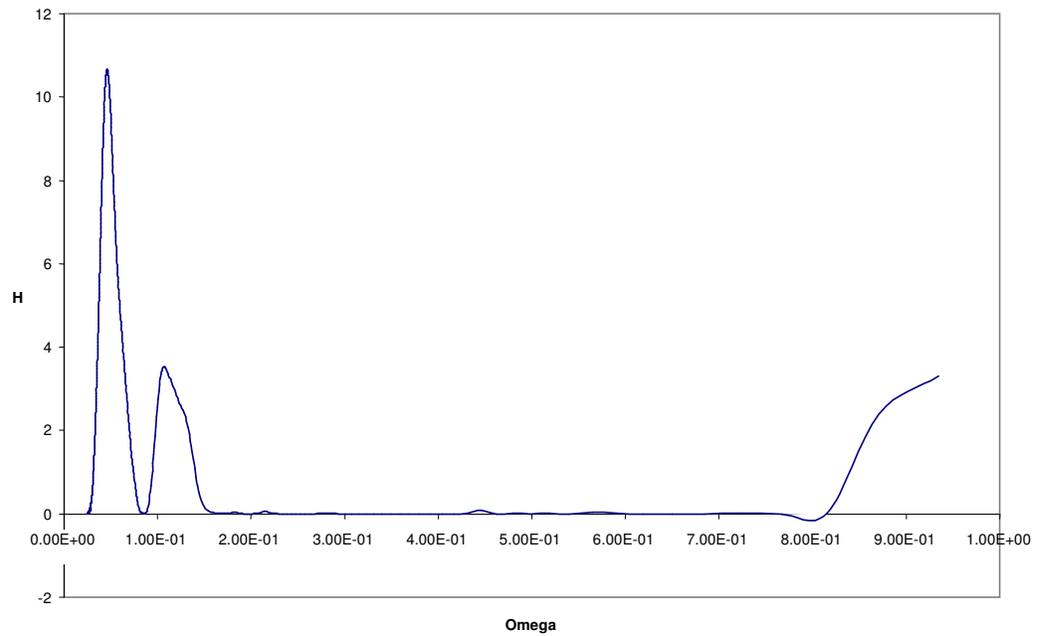

**Fig. 4:** P-ω graph obtained by Periodogram method of 5-day-long samples from Super-Kamiokande-I detector during the period from June, 1996 to July, 2001

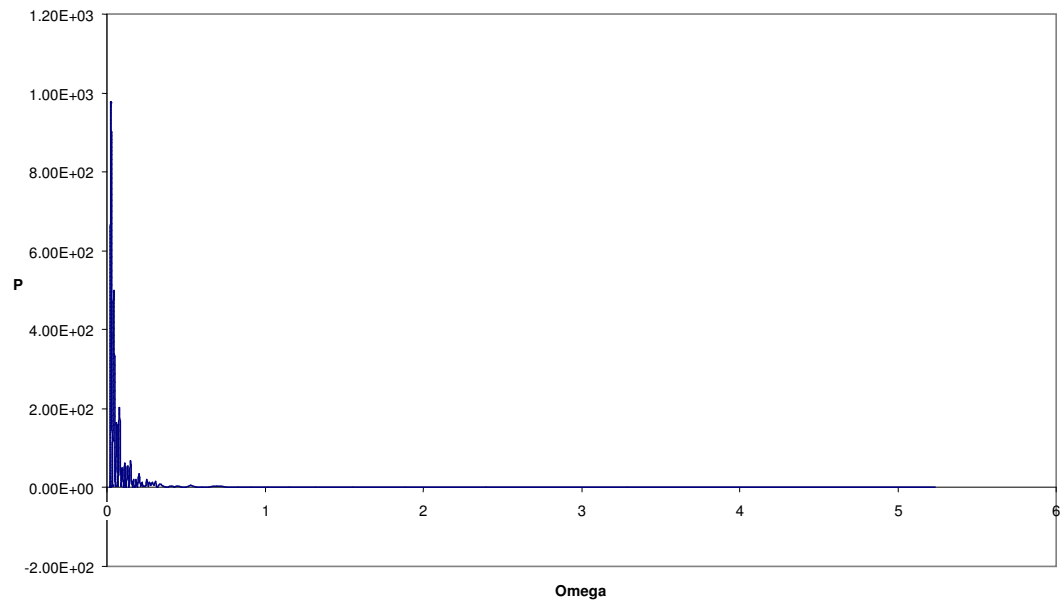

**Fig. 5:** P-ω graph obtained by Periodogram method of 10-day-long samples from Super-Kamiokande-I detector during the period from June, 1996 to July, 2001.

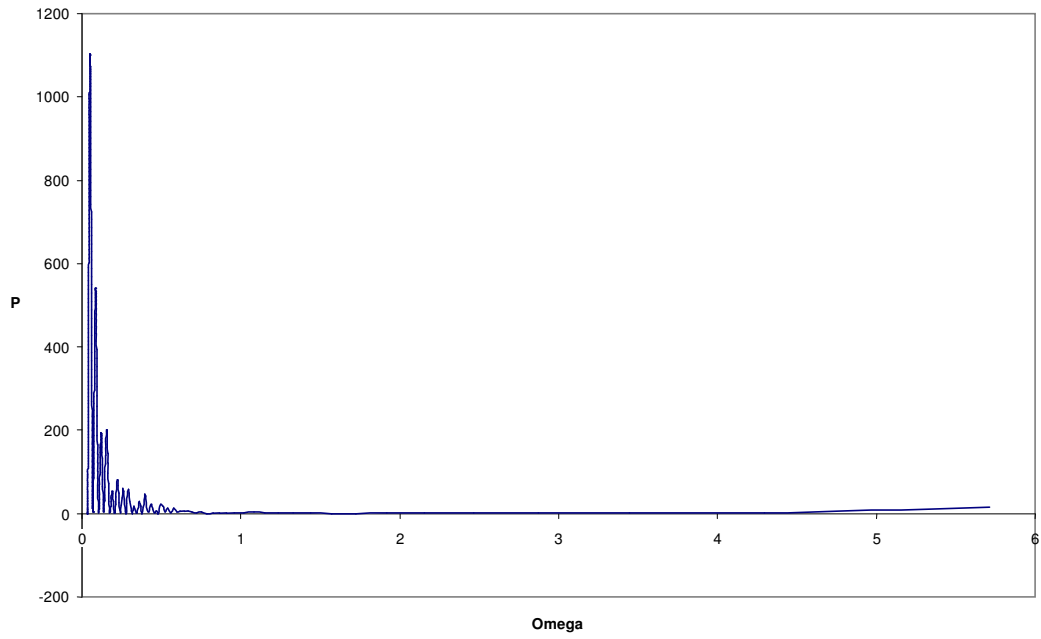

**Fig. 6:** P-ω graph obtained by Periodogram method of 45-day-long samples from Super-Kamiokande-I detector during the period from June, 1996 to July, 2001.

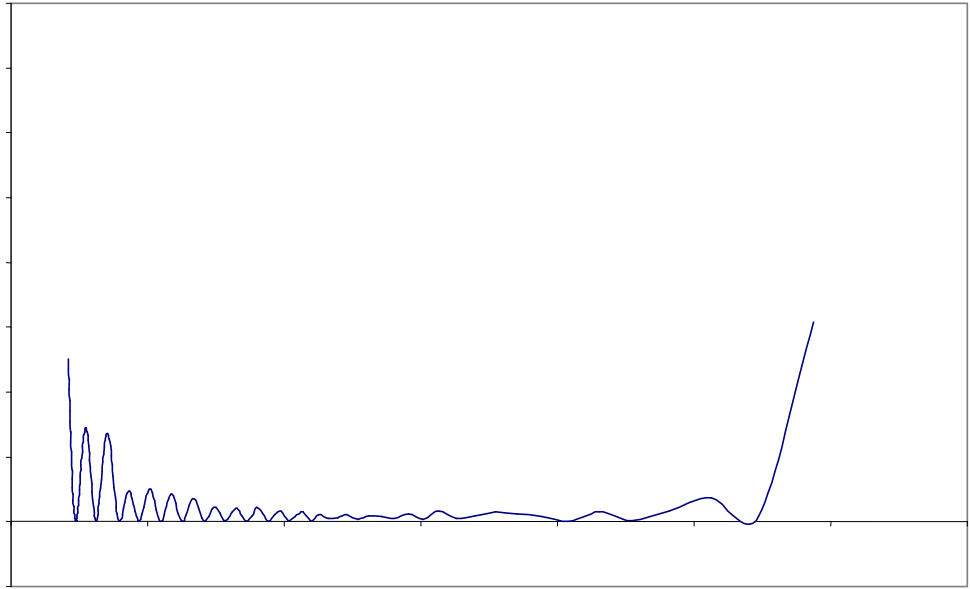

## V. Discussion:

Comparing the results obtained in Ferraz-Mello method and Scargle method we can confirmly say that (1) shows periodicities around 0.21 – 0.22, 1.15 – 1.98, 0.67 – 0.77, 6.72 – 6.95, 12.05 – 13.24, 22.48 – 24.02 months, (2) gives periodicities around 0.39 – 0.45, 1.31 – 2.23, 5.20-5.32, 9.43 – 11.65, 13.54 – 14.38 months and for (3) we are not getting any periodicity in months which can be compatible to both the methods. The observed periods of around 0.45 months and 1.31 months in (2) are not appreciably different from the periods of 13.75 days (at the frequency 26.57 $y^{-1}$) and 38.75 days (at the frequency 9.42 $y^{-1}$) respectively obtained by Caldwell and Sturrock[10]. But here for these two obtained periods the significance level is not so high (around 88%). But our obtained significance level 88.24% for 0.45 months periodicity in (2) is higher compared to the significance level 81.70% for similar periodicity of 13.76 days obtained by Yoo *et a*.[9].

## VI.	Solar Neutrino Flux of S.K.-I in the Solar Activity Cycle:

It has been suggested that [16, 17] there exist five phases in many solar activities i.e. sunspot numbers, solar magnetic fields and solar neutrino flux data in the Homestake, SAGE and GALLEX detector. S.K.-I data starts from 31 May 1996 to 15 July 2001. In that period both sunspot minimum and sunspot maximum fall. To see this we have taken the six month average of S.K. data. The following table shows the six month average data:

| Year | January to June | July to December |
|------|-----------------|------------------|
| 1996 |                 | 2.289            |
| 1997 | 2.326           | 2.282            |
| 1998 | 2.356           | 2.468            |
| 1999 | 2.489           | 2.425            |
| 2000 | 2.295           | 2.297            |
| 2001 | 2.494           |                  |

We take the solar neutrino flux 2.289 at the period from July to December 1996 as Phase I, 2.468 ~ 2.489 at the period from July to December 1998 to January to June 1999 as Phase II and 2.295 ~ 2.297 at the period from January to June 2000 to July to December 2000 as Phase III and 2.494 at the period from January to June 2001 as phase IV and Phase V will occur after 2-3 years from July 2001. These four phases coincide with the phases described for solar neutrino flux data from Homestake, GALLEX, SAGE and solar magnetic field and sunspot data during the 23$^{rd}$ sunspot cycle.

It should be noted that common five phases of neutrino flux, general magnetic field, sunspot numbers etc. in the solar activity cycle probably indicate that they are all related but the exact transfer of energy to produce such a relation cannot yet be formulated.